\documentstyle[11pt,paspconf,epsf]{article}

\begin{document}

\title{COBE Observations of Interstellar Line Emission}

\author{E. L. Wright}
\affil{UCLA Physics \& Astronomy Department,
    Los Angeles, CA 90095-1562}

\begin{abstract}
The FIRAS instrument on the {\sl COBE} satellite has conducted an unbiased
survey of the far-infrared emission from our Galaxy.  The first
results of this survey were reported by Wright {\it et al.} (1991),
and later conclusions were reported in Bennett {\it et al.} (1994).
I report the results of analyses of this
spectral survey, which includes emission lines from
158~$\mu$m C$^+$, 122~$\mu$m and 205~$\mu$m N$^+$,
370~$\mu$m and 609~$\mu$m C$^0$, and CO J=2-1 through 5-4.
The morphological distribution along the galactic
plane ($b=0^\circ$) for all the spectral line
emission, and the high galactic latitude intensities of the strong
C$^+$ and 205~$\mu$m N$^+$ emission are discussed.
The high galactic latitude intensity of the 158 \micron\
fine structure transition from C$^+$ is $({\rm C}^+\; 158\;\mu$m$)
\approx (1.43\pm 0.12)\times 10^{-6}\csc{\vert b\vert}\; 
$erg$\;$cm$^{-2}\;$s$^{-1}\;$sr$^{-1}$ for $\vert b\vert > 15^\circ$,
and it decreases more rapidly than the far infrared intensity
with increasing galactic latitude.
\end{abstract}


\section{Introduction}

The interstellar medium is heated primarily by photons interacting
with grains, producing hot electrons by the photoelectric effect
(Watson 1972).
Other heating is supplied by the excess energy given to electrons
detached from atoms during photoionization, and by the excess energy
given to atoms during the photodissociation of molecules.  The hot
electrons and atoms collide with the thermal particles and add
energy.  To maintain a steady state, this added energy must be
radiated by the gas.  The temperature of the gas will increase 
until the radiation process with lowest excitation threshold
becomes effective.  In the neutral atomic interstellar medium,
the fine structure line of the C$^+$ ground state with an
excitation energy corresponding to 91~K is the first effective
cooling channel to open, so the C$^+$ line at 157.7~$\mu$m is
expected to be strong (Dalgarno \& McCray 1972).
The ionization threshold for C is 11.26~eV,
so it will be ionized even when H is neutral.  O and N have higher
ionization thresholds and will be neutral in H~I regions.
The ground electronic state of O is an inverted fine structure 
triplet, so the first transition from the ground state corresponds
to the 63~$\mu$m line, and thus requires a considerably higher 
temperature for excitation.

In the warm ionized medium, the CNO elements will all be singly
ionized.  The cooling channel with the lowest excitation energy
will be the N$^+$ line at 205~$\mu$m, but the 158~$\mu$m line
of C$^+$ will still be strong.  With a typical temperature of
a few thousand degrees in the warm ionized medium, there will
be many more channels for cooling, so none of these lines will
dominate the cooling in the way the C$^+$ line dominates in 
H~I regions.

In molecular gas, the rotational transitions of the CO molecule
provide low threshold energy cooling channels.  The density and
radiation field in regions with CO favor the presence of neutral
C instead of C$^+$.  The ground electronic state C$^0$ is a fine
structure triplet with lines at 609 and 370~$\mu$m.

In this paper I will discuss the observations of these lines
by the FIRAS instrument on the {\sl COBE} satellite.  
These data were first discussed by Wright {\it et al.} (1991) 
which included the first detection ever of the 205~$\mu$m N$^+$
line.  Bennett {\it et al.} (1994) discuss the distribution of
these lines around the sky.

\section{Observations}

\begin{figure}
\plotone{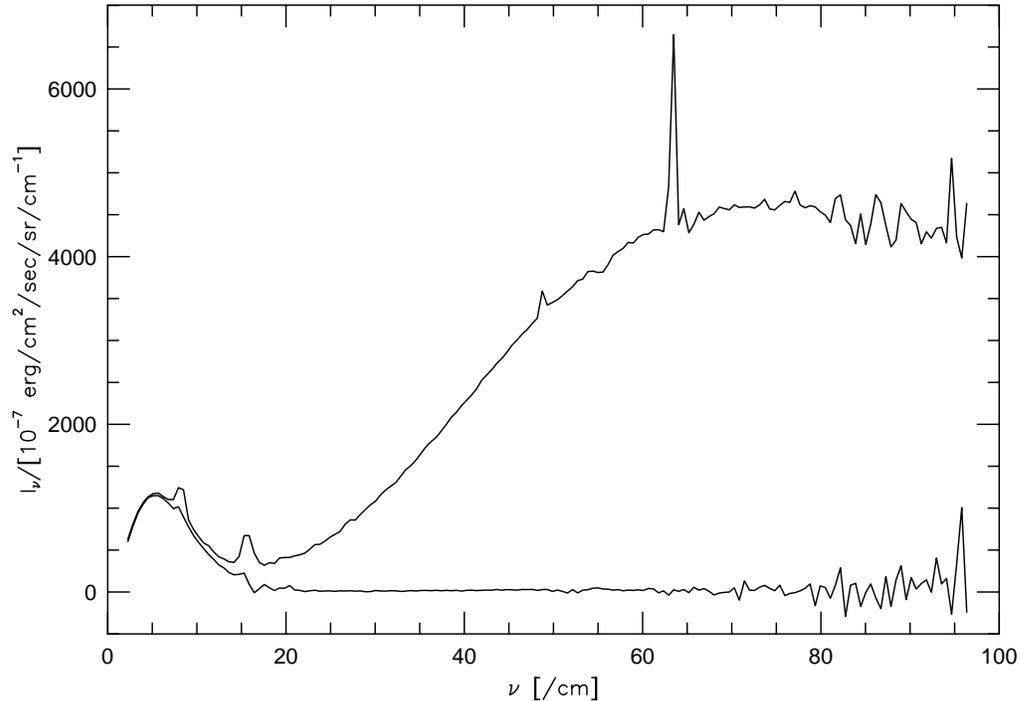}
\caption{FIRAS Right High Spectra of the Galactic Center
and the Lockman Hole.\label{gc_lh.eps}}
\end{figure}

The FIRAS instrument is a polarizing Michelson interferometer
first described by Martin \& Puplett (1970).  Using free-standing
wire grids as polarizers allows a single instrument to measure
many octaves of spectrum at once.  Figure \ref{gc_lh.eps} shows
spectra of the Galactic Center pixel and the pixel with the lowest
$N_H$ (Lockman, Jahoda and McCammon 1986)
taken from the Right High channel of FIRAS.  
The high frequency part of the spectrum ($> 20$~cm$^{-1}$)
was separated from the cosmic signal by a dichroic beamsplitter, but
the residual reflectance of this filter allows one to see most of
cosmic blackbody spectrum as well.

The FIRAS instrument uses bolometric detectors, and is thus
approximately equally sensitive to lines at any frequency.
However, variations of filter transmission and modulation
efficiency modify this equality and must be determined using
calibration observations.  The in-flight calibrators are
essentially blackbodies, while the galactic spectrum is a
very dilute modified Planck function.
Thus the 50 part per million precision
achieved in the search for deviations of the CMB from a blackbody
(Fixsen {\it et al.} 1996)
does not imply 50 part per million precision in the galactic
dust and line fluxes.  In particular, small ripples in the gain
calibration could be confused with weak line fluxes in regions
where the continuum is high.

\begin{figure}
\plotone{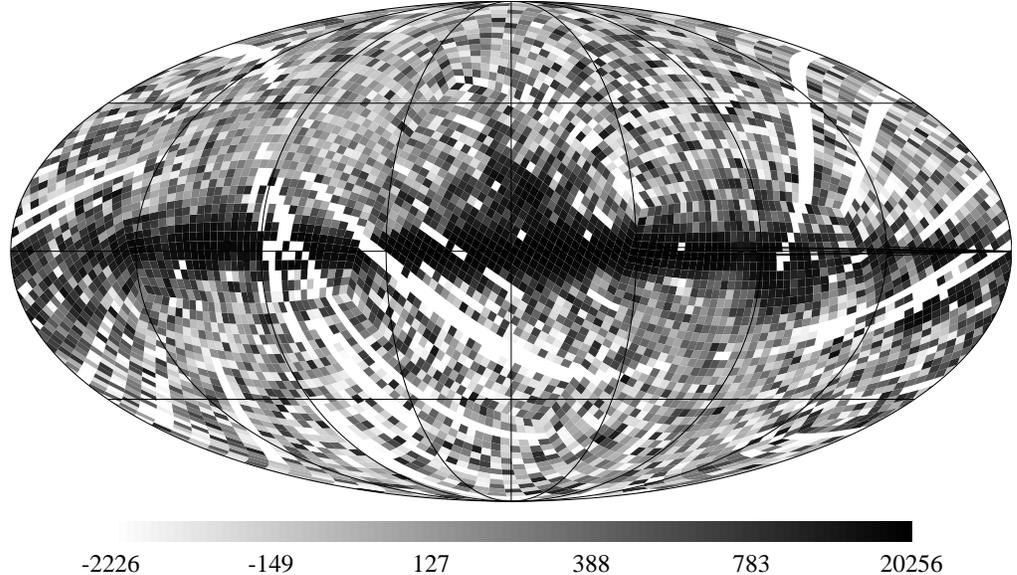}
\caption{FIRAS C$^+$ line map in units of 
$10^{-8}$~erg/cm$^2$/sec/sr/cm$^{-1}$.\label{cplus3.eps}}
\end{figure}

The bright C$^+$ and N$^+$ lines are easily visible in the spectrum
of the galactic center, and by fitting a line profile plus baseline
to the spectrum of each pixel, maps can be generated.
Figure \ref{cplus3.eps} shows a map of the C$^+$ line flux.

\begin{figure}
\plotone{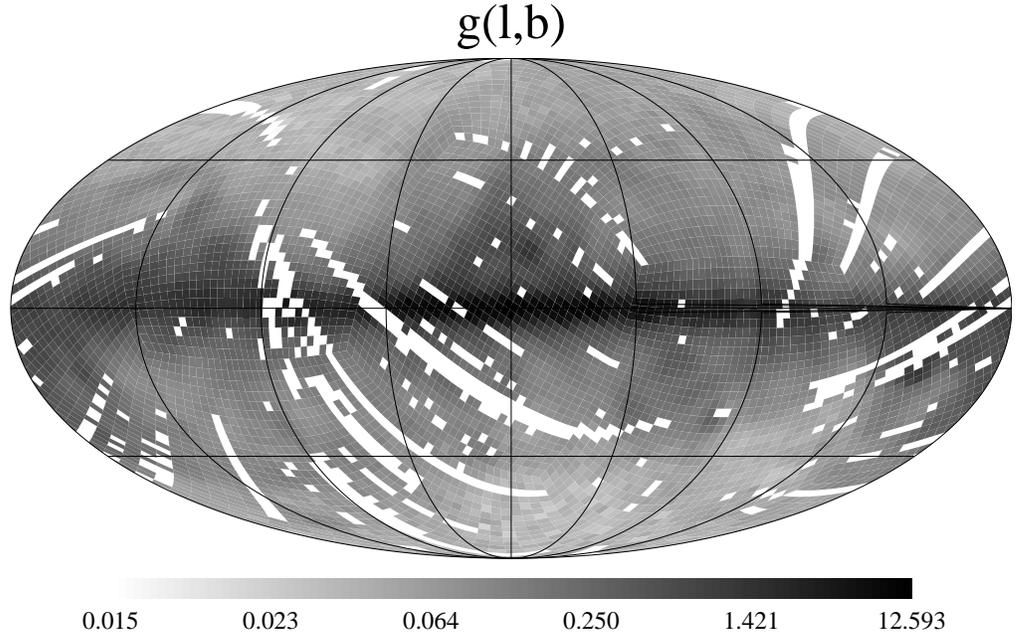}
\caption{FIRAS dust map $g(l,b)$.\label{dust3.eps}}
\end{figure}

Higher SNR on the galactic signal can be obtained by
averaging all of the pixels together.  But an arbitrary
isotropic signal must be included in the fit, as well
as the dipole variation of the CMB temperature.  The
following approach was used:
\begin{enumerate}
\item
Make maps of the cosmic temperature variation and the galactic
dust distribution using an initial guess for the mean galactic spectrum,
$G_\circ(\nu)$:
$$I_\nu(l,b) \approx B_\nu(T_\circ+\Delta T(l,b)) + g(l,b)
G_\circ(\nu)$$

\item
Do a spectral fit to derive the isotropic cosmic spectrum, $I_\circ(\nu)$; 
the spectrum of the dipole, $D(\nu)$; and the mean spectrum of the galaxy, 
$G(\nu)$:
$$
I_\nu(l,b) \approx I_\circ(\nu) + D(\nu) \cos \theta + G(\nu) g(l,b)
$$
where $\theta$ is the angle between the LOS $(l,b)$ and the hot pole
of the dipole.

\item
Iterate using $G(\nu)$ instead of $G_\circ(\nu)$.

\end{enumerate}
This procedure can be applied to a subset of the sky as well.
Note that changes in the isotropic spectrum have no effect
of the derived $G(\nu)$, which is determined solely from the
variations of the spectrum with position on the sky.

Wright {\it et al.} (1991) applied this procedure to the early
data from FIRAS, and obtained a map of $g(l,b)$ over part of the
sky and a spectrum $G(\nu)$ which was analyzed for line strengths
and dust emission parameters.  Figure \ref{dust3.eps} shows 
$g(l,b)$ for the entire FIRAS data set.  Note that the polar flux
(slope with respect to $\csc\vert b\vert$) is 0.05 units on this
map.

\begin{figure}[htb]
\plotone{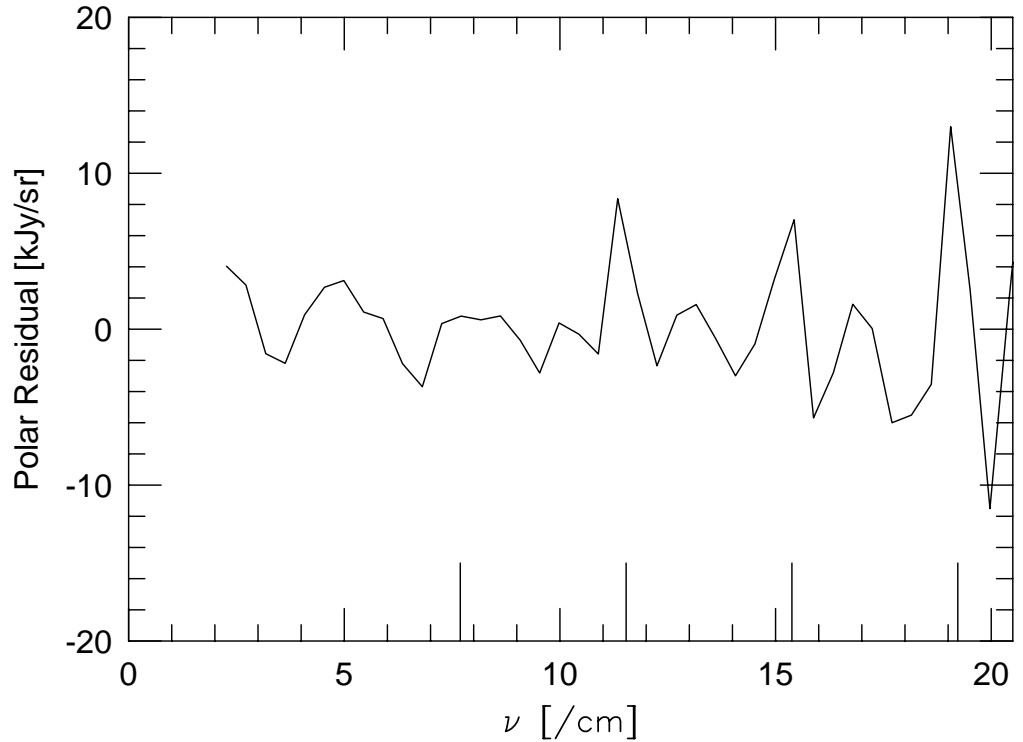}
\caption{Fixsen {\it et al.} (1996) $G(\nu)$ minus a
modified blackbody. The positions of the 2-1 through 5-4 CO lines
are marked.
\label{g_resids.eps}}
\end{figure}

The spectrum $G(\nu)$ derived only from data taken away from the
galactic plane is given by Fixsen {\it et al.} (1996).  Figure
\ref{g_resids.eps} shows this $G(\nu)$ after a modified Planck
function is subtracted.  The 3-2 and 4-3 CO lines have about the
same strength in this spectrum as they do in the all-sky 
$G(\nu)$, while the 5-4 line is stronger.  However, the SNR of
this spectrum is quite low since all of the strong galactic
signal has been ignored.

\section{CO Analysis}

\begin{figure}[t]
\plotone{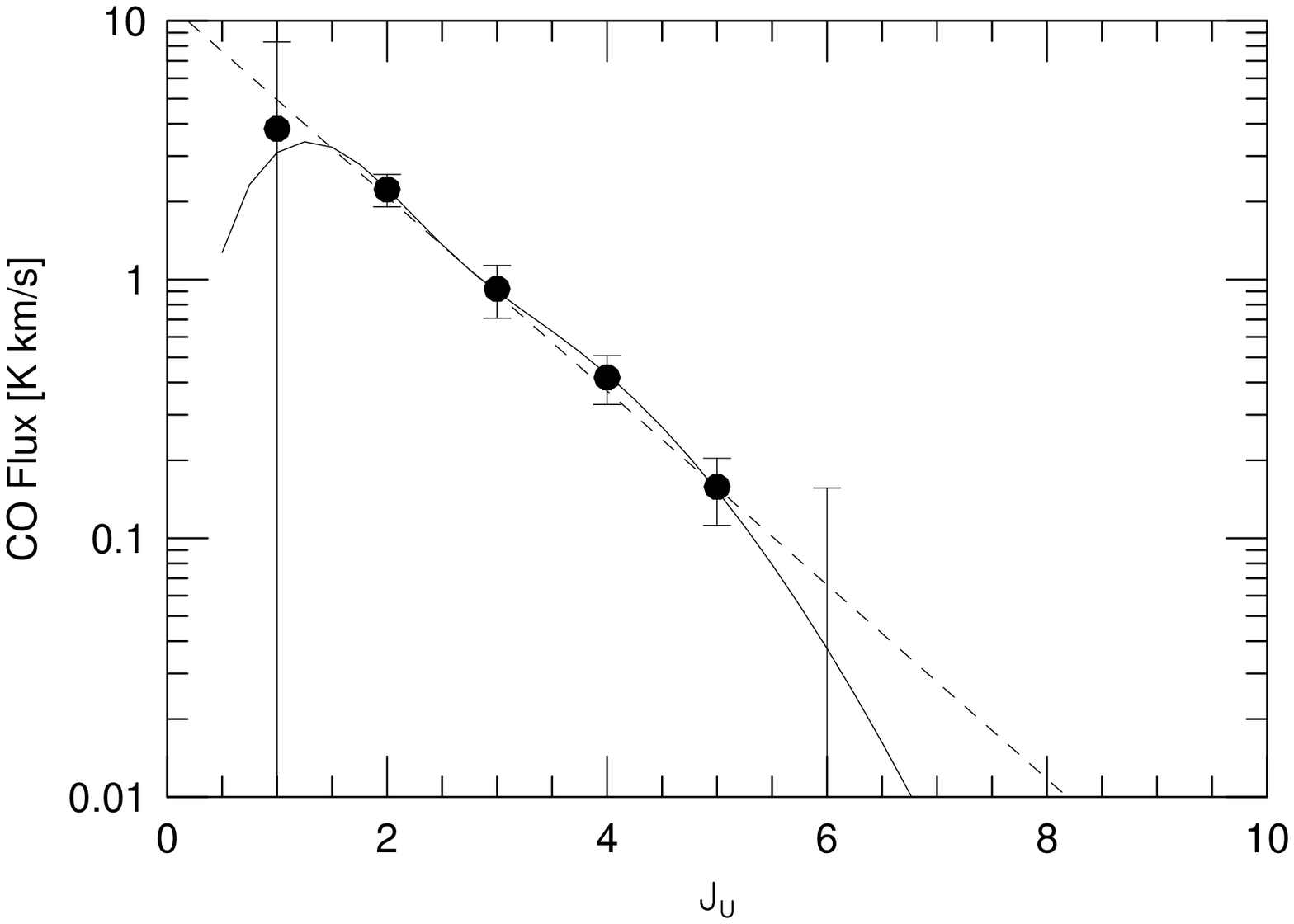}
\caption{FIRAS CO line strengths at $g(l,b)=1$.
The solid line shows a two-temperature model fit.\label{co_fit.eps}}
\end{figure}

FIRAS detected the CO lines from 2-1 to 5-4.  In terms of power
in erg/sec, the 5-4 line is quite strong.  The excitation of CO
to the $J=5$ level which has a radiative lifetime of only one day
would seem to be quite difficult in the high galactic latitude
molecular clouds seen by Magnani, Blitz and Mundy (1985).
This excitation looks more reasonable if the CO line strengths
are plotted in units of K-km/sec, as shown as in Figure
\ref{co_fit.eps}.  The almost perfect fit of these line
strengths to $\exp(-bJ_U)$ (dashed line) suggests a thermal 
distribution, but since the energy of the CO levels is
$\propto J(J+1)$ a mixture of temperatures is needed:
the solid curve shows a model of optically thin CO
with 88\% at 5~K and 12\% at 19~K.  While this gives a good fit,
the number of parameters equals the number of detected lines,
so there are no remaining degrees of freedom.  This two
temperature model predicts a CO 1-0 line flux of 0.15 K-km/sec
at the galactic pole.  The standard ratio of CO 1-0 line
flux to molecular hydrogen column density,
$N(H_2)/I(CO) = 3 \times 10^{20}/($cm$^2\;$K$\;$km/sec$)$
(Young \& Scoville 1991), gives
$N(H_2) = 5 \times 10^{19}$ so this molecular gas is about
1/3 of the atomic gas at high latitudes.
Blitz, Bazell \& Desert (1990) estimate that the mass of the high
latitude molecular clouds is 33\% of the neutral atomic mass,
implying that the FIRAS CO flux is due to the smeared out
emission from the MBM clouds.
The high latitude CO clouds have lower densities and lower
extinctions than most molecular clouds, so a fluorescent
cascade following vibrational or electronic excitation could 
provide most of the high-$J$ CO.

\section{Conclusion}

The FIRAS data on the Milky Way show the primary cooling
mechanisms in the ISM.  99.7\% of the power is radiated
by dust heated directly by the interstellar radiation
field.  Gas in the ISM radiates 0.3\% of the power in
the C$^+$ line at 158~$\mu$m.  The next most luminous
species seen by FIRAS is N$^+$, with 0.03\% of the power
in the 205~$\mu$m line.  C$^0$ and CO each radiate about
0.003\% of the total power.

\acknowledgments

This work was made possible by the
dedication of hundreds of scientists and engineers who worked
to make the {\sl COBE} project a big success.

% That's the end of the main body of the paper.  Now we will have some
% back matter.

\begin{question}{Dr.\ Nakagawa}
What is the temperature range derived from the CO lines?
\end{question}
\begin{answer}{Prof.\ Wright}
The fit gives 5 K and 19 K.
\end{answer}

\begin{question}{Prof.\ Greenberg}
When you say 88\% of the CO at 5 K and
12\% at 19 K do you imply that the CO is thermally
coupled to the dust?  I find this difficult to believe. 
Is this just a coincidence?
\end{question}
\begin{answer}{Prof.\ Wright}
I hope so!  Collisional heat transfer between gas and dust
is quite slow.
\end{answer}

\begin{question}{Dr.\ Taniguchi}
Your continuum and [C~II] maps show an emission 
component North of the Galactic Center.  What is it?
\end{question}
\begin{answer}{Prof.\ Wright}
That is the Sco-Oph dark cloud complex.
\end{answer}

\begin{question}{Dr.\ Taniguchi}
Is the northern extension at $l=0$ and the
southern extension at the anti-center due to warping of
the disk?
\end{question}
\begin{answer}{Prof.\ Wright}
No, those are both very nearby cloud complexes. 
The warp of the disk is a much smaller angle.
\end{answer}

\begin{question}{Dr.\ Giard}
What is the upper limit on the ortho:para
ratio in H$_2$O based on the apparent absorption at the
Galactic Center?
\end{question}
\begin{answer}{Prof.\ Wright}
First, we do not claim the absorption at $l=0$ is
real.  But if it is, the line strengths shown
in Figure 2 of Bennett {\it et al.} (1994) imply 
an equivalent width of $W_p = (3\pm 4)\times 10^{-3}$~cm$^{-1}$ for the
para line at 269.3~$\mu$m
while the ortho line at 538.3~$\mu$m has 
$W_o = (0\pm 6)\times 10^{-3}$~cm$^{-1}$.
A given equivalent width in the ortho line implies 4 times more column 
density than the same equivalent width in the para line because the
frequency is half as large and the transition matrix element
$|\mu_{ij}|^2$ is half as large.
Thus ortho:para column density ratios ranging from 0 to more than 3:1 are 
all consistent with these equivalent widths.
If both lines are optically thick, as seems likely, then $W_o$ should
be one half of $W_p$ because of the frequency factor, which is also
consistent with the observations.  In the optically thick case we
have no information on the ortho:para ratio.
\end{answer}

\begin{question}{Dr.\ Dwek}
How do you interpret the two-temperature dust
fit in terms of a physical dust model?
\end{question}
\begin{answer}{Prof.\ Wright}
It could be caused by a shoulder in the dust
emissivity law.  This would explain the constancy of the
cold dust temperature and the cold/warm opacity ratio.
\end{answer}

\end{document}